\documentstyle[stwol,epsf]{article}

\newcommand{\AmS}{{\protect\the\textfont2
  A\kern-.1667em\lower.5ex\hbox{M}\kern-.125emS}}
\newcommand{\figV}      {mssmbsg1.6_11_sw_znn.eps}    
\newcommand{\sintw}     {sintw.eps}                   
\newcommand{\alphas}  {alphas.eps}                    

\newcommand{\bq}{\begin{equation}}
\newcommand{\eq}{\end{equation}}
\newcommand{\beq}  {\begin{eqnarray}}
\newcommand{\eeq}  {\end{eqnarray}}
\newcommand{\rG}   {{\rm GUT}}
\newcommand{\MG}   {{\ifmmode M_\rG         \else $M_\rG$          \fi}}
\newcommand{\mb}   {{\ifmmode m_{b}         \else $m_{b}$          \fi}}
\newcommand{\mt}   {{\ifmmode m_{t}         \else $m_{t}$          \fi}}
\newcommand{\agut} {{\ifmmode \alpha_\rG    \else $\alpha_\rG$     \fi}}
\newcommand{\mgut} {{\ifmmode M_\rG         \else $M_\rG$          \fi}}
\newcommand{\mze}  {{\ifmmode m_0           \else $m_0$            \fi}}
\newcommand{\mha}  {{\ifmmode m_{1/2}       \else $m_{1/2}$        \fi}}
\newcommand{\tb}   {{\ifmmode \tan\beta     \else $\tan\beta$      \fi}}
\newcommand{\mz}   {{\ifmmode M_{Z}         \else $M_{Z}$          \fi}}
\newcommand{\ai}   {{\ifmmode \alpha_i      \else $\alpha_i$       \fi}}
\newcommand{\aii}  {{\ifmmode \alpha_i^{-1} \else $\alpha_i^{-1}$  \fi}}
\newcommand{\MSb}  {{\ifmmode \overline{\rm MS} \else
                             $\overline{\rm MS}$                   \fi}}
\newcommand{\DRb}  {{\ifmmode \overline{\rm DR} \else
                             $\overline{\rm DR}$                   \fi}}
\newcommand{\DRbar}{{\ifmmode \overline{DR} \else $ \overline{DR}$ \fi}}
\newcommand{\msusy}{{\ifmmode M_{SUSY}      \else $M_{SUSY}$       \fi}}
\newcommand{\as}   {{\ifmmode \alpha_s      \else $\alpha_s$       \fi}}
\newcommand{\asmz} {{\ifmmode \alpha_s(M_Z) \else $\alpha_s(M_Z)$  \fi}}
\newcommand{\tal}  {{\ifmmode \tilde{\alpha} \else $\tilde{\alpha}$ \fi}}

\newcommand{\sws}  {{\ifmmode \;\sin^2\theta_W
                     \else    $\;\sin^{2}\theta_{W}$               \fi}}
\newcommand{\cws}  {{\ifmmode \;\cos^2\theta_W  
                     \else    $\;\cos^{2}\theta_{W}$               \fi}}
\newcommand{\sw}   {{\ifmmode\;\sin\theta_W\else $\sin\theta_{W}$  \fi}}
\newcommand{\cw}   {{\ifmmode\;\cos\theta_W\else $\;\cos\theta_{W}$\fi}}
\newcommand{\tw}   {{\ifmmode\;\tan\theta_W\else $\;\tan\theta_{W}$\fi}}
\newcommand{\bsg}  {{\ifmmode \b\rightarrow s\gamma
\else $b\rightarrow s\gamma$ \fi}}
\newcommand{\Bbsg}  {{\ifmmode BR(\b\rightarrow s\gamma)
\else $BR(b\rightarrow s\gamma)$ \fi}}

\newcommand{\nn}   {\nonumber \\}

\def\be{\begin{equation}}
\def\ee{\end{equation}}
\def\bea{\begin{eqnarray}}
\def\eea{\end{eqnarray}}

\begin{document}
\title{Global Fits of the MSSM and SM to Electroweak Precision Data}

\author{W. DE BOER }

\address{Inst. f\"ur Experimentelle Kernphysik, Univ. of Karlsruhe, \\ 
        Postfach 6980, 76128 Karlsruhe, Germany}


\twocolumn[\maketitle\abstracts{The Minimal supersymmetric extension 
of the Standard Model (MSSM) with light stops,
charginos or pseudoscalar Higgs bosons has been
suggested as an explanation of the too high value of the 
branching ratio of the $Z^0$ boson into $b$~quarks ($R_b$ anomaly).
A program including all radiative corrections to the MSSM at the same
level as the radiative corrections to the SM has been developed and used
to perform global fits to all electroweak data from LEP, SLC and the Tevatron.
Recent updates on electroweak data,  presented at 
this conference, reduce the $R_b$ anomaly from a 
3.2$\sigma$ to a $1.8 \sigma$ effect.
In addition, the $b\rightarrow s\gamma$ decay is $30\%$  below the SM prediction.
In the MSSM light stops and light  charginos increase $R_b$ and decrease the  
$b\rightarrow s\gamma$ rate, so both observations can be brought into 
agreement with the MSSM for the same region of parameter space.
However, the resulting    $\chi^2$ value for the MSSM fits is 
only marginally lower. In addition,  the splitting in the stop 
sector has to be  unnaturally high, so it remains to be seen if these effects
are real or due to a fluctuation.}]
\section{Introduction}
 Previous  LEP data showed a too high value of $R_b$ (3.2$\sigma$) and a too low
value of $R_c$ (2$\sigma$) where $R_{b(c)}$ is the ratio
$R_{b(c)}=\Gamma_{Z^0\rightarrow b\bar b (c\bar c)}/\Gamma_{Z^0\rightarrow q\bar q}$.
In the past it has been shown by several groups that it is possible to increase $R_b$
using minimal supersymmetric models (MSSM) with light charginos, stops or Higgses, which yield
positive contributions to the $Zb\bar b$ vertex, although no improvement in $R_c$ could be
obtained\cite{yel1}\nocite{boufi,chan2,ell1,garc3,kan1,kan2,garcia}-\cite{garcia2}.
Recent updates of electroweak data \cite{Lep3} show no significant deviation of $R_c$ from
the Standard Model (SM) prediction, and a value of 
$R_b$ which is $1.8\sigma$ above the SM value. 
Such a moderate deviation in $R_b$ can  either be  a fluctuation 
or may originate from genuine MSSM contributions.
In this paper an equivalent analysis of all electroweak data,
both in the SM and its supersymmetric extension, is described
using all actual electroweak data from Tevatron, 
LEP and SLC \cite{Lep3},
the measurement of
$\frac{BR(b\rightarrow s\gamma)}{BR(b\rightarrow ce\bar\nu)}$ 
from CLEO \cite{cleo}
and limits on the masses of supersymmetric
particles.
Further details of the  procedure and extensive references are given
  elsewhere\cite{hollik1}.
\section{Results}
\subsection{\it Standard Model Fits}
\begin{figure}
 \begin{center}
  \leavevmode
  \epsfxsize=7.cm
  \epsffile{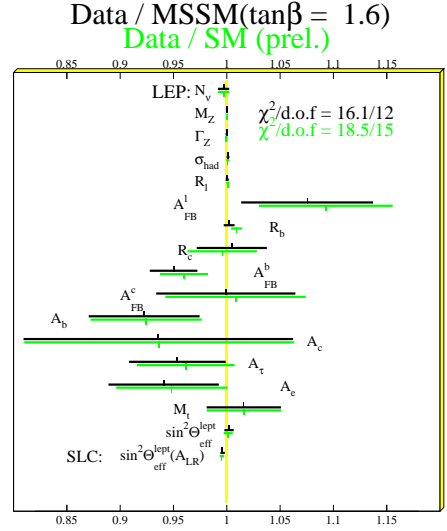}
 \end{center}
\caption{\label{\figV}
Fit results normalized to the SM- and MSSM ($\tan\beta=1.6$) values.
The difference in $\chi^2/d.o.f$ between the SM and MSSM originates 
mainly from $R_b$.}
\end{figure}

The SM cross sections are determined by $M_Z,m_t,m_h,G_F,\alpha,\alpha_s$.
From the combined CDF and D0 data $m_t$ has  been determined to be $175\pm6$ GeV\cite{Lep3},
so the parameters with the largest uncertainties are $m_h$ and $\alpha_s$.
The error on the finestructure constant $\alpha$ is limited by the uncertainty in the
hadronic cross section in $e^+e^-$ annihilation at low energies, 
which is used to determine the vacuum polarization contributions to $\alpha$.
The error was taken into account by considering $\alpha$ to be a free parameter in the fit
and constraining it to the value $1/\alpha=128.89\pm0.09$\cite{jegerlehner}.
If this error is not taken into account, the error on the Higgs mass 
is underestimated by 30\%. 
Using the input values discussed in the introduction yields:
\begin{eqnarray*}
\alpha_s=0.120\pm0.003\\
m_t=172.0^{+5.8}_{-5.7}~{\rm GeV}\\
m_H=141^{+140}_{-77}~{\rm GeV} \\
sin^2\theta_{\overline{MS}}=0.2316\pm0.0004.
\end{eqnarray*}
Minor deviations from the EWWG fit results\cite{Lep3}
 are due to the incorporation
of the $b\rightarrow \gamma$ data from CLEO\cite{cleo},
 which are  important for
the MSSM fits described below.
The value of $sin^2\theta_{\overline{MS}}$ is within errors  equal to 
$\sin^2\Theta_{eff}^{lept}$.
The main contributions to the
$\chi^2/d.o.f=19.6/15$ originate from $\sin^2\Theta_{eff}^{lept}$ from SLD  
($\Delta\chi^2=4.9$), $R_b$ ($\Delta\chi^2=3.1$) 
and $A_{FB}^b$  ($\Delta\chi^2=3.5$). The SM agreement is good for all 
other observables, including $R_c$, as shown in fig. \ref{\figV}.

\begin{figure}
 \begin{center}
  \leavevmode
  \epsfxsize=7cm
  \epsffile{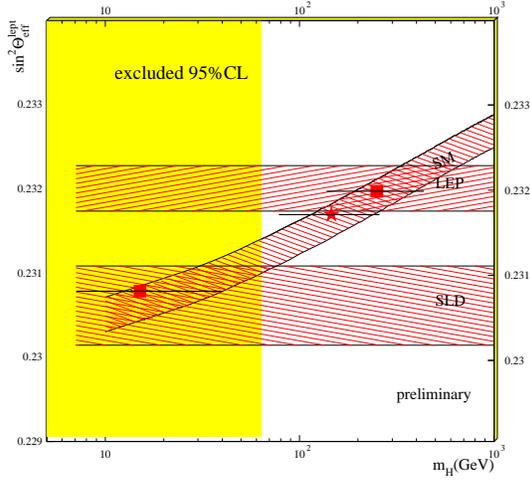}
 \end{center}
\caption{\label{\sintw}
Dependence of the SM $\sin^2\Theta_{eff}^{lept}$ on the Higgs mass. The top
mass $m_t=175\pm 6$~GeV was varied within its error, as shown by the dashed band
labelled SM. 
The SLD and the LEP measurement of  $\sin^2\Theta_{eff}^{lept}$ 
are also shown as horizontal bands.
The SLD value yields a Higgs mass below the recents limits 
 by direct Higgs searches at LEP
(shaded area).
} 
\end{figure}

\begin{figure}
 \begin{center}
  \leavevmode
  \epsfxsize=7cm
  \epsffile{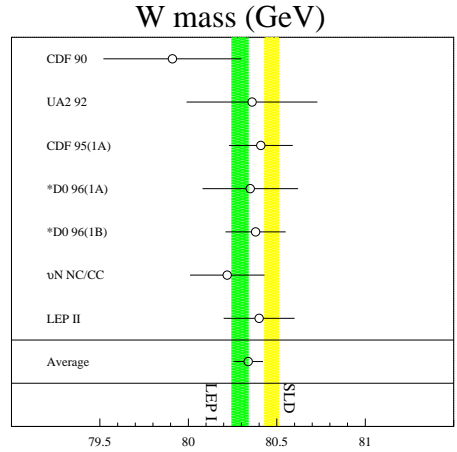}
 \end{center}
\caption[]{\label{mw}
A compilation of the W-masses. The vertical lines indicate the predictions
from the LEP I and   SLD electroweak data determining 
$\sin^2\Theta_{eff}^{lept}$ and the data points represent the various direct measurements of $M_W$. The preliminary LEP II value was taken from 
presented at the LEPC Meeting\cite{opalguy}.
} 
\end{figure}

The  dependence of $\sin^2\Theta_{eff}^{lept}$ on the SM Higgs mass 
is approximately logarithmic
(see fig. \ref{\sintw}).
The LEP data alone without SLD yield $m_H\approx240$~GeV, 
while $\sin^2\Theta_{eff}^{lept}$
from SLD corresponds to $m_H\approx15$~GeV,
as indicated by the squares in fig. \ref{\sintw}.
The latter value is excluded by the lower limit of 58.4~GeV from 
the combined LEP experiments~\cite{rev96}.
The different values of $\sin^2\Theta_{eff}^{lept}$ 
from LEP and SLD translate into different
predictions for $M_W$, as shown in fig. \ref{mw}. 
The present $M_W$ measurements,
including the preliminary value from the LEP II measurements at 
161 GeV\cite{opalguy},  
lie in between these  predictions.
\begin{figure}
 \begin{center}
  \leavevmode
  \epsfxsize=7cm
  \epsffile{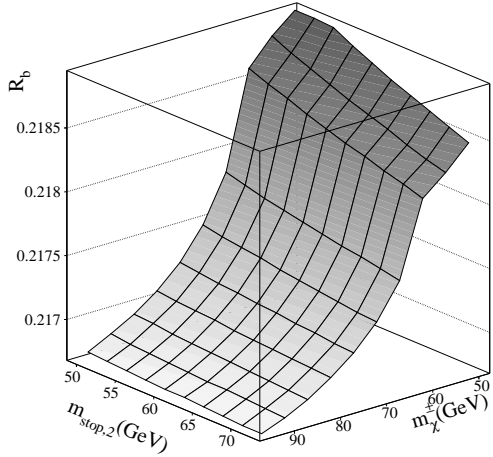}
 \end{center}
\caption[]{\label{rb}
$R_b$ as function of the stop and chargino masses.
 The experimental value $R_b=0.2178\pm 0.0011$ is clearly 
above the SM value of 0.2158 and can be obtained for light charginos and stops.
}
\end{figure}
\begin{figure}
 \begin{center}
  \leavevmode
  \epsfxsize=7cm
  \epsffile{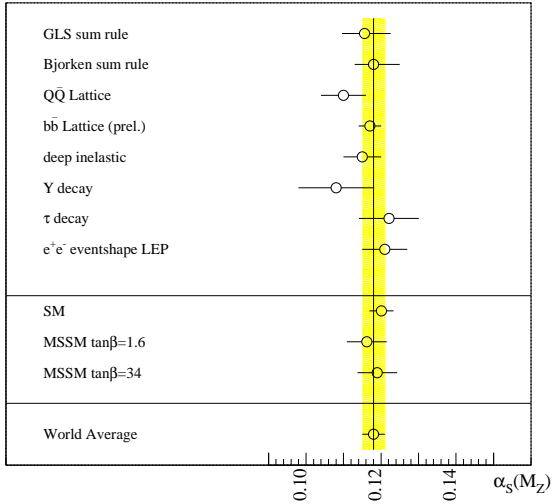}
 \end{center}\vspace{-0.8cm}
\caption[]{\label{\alphas}
Comparison  of $\alpha_s$ determinations with the fit results.
The data have been taken from \protect\cite{rev96} 
and \protect\cite{schmelling}.
 Note the remarkable agreement between the new precise determinations:
the  values from the SM and MSSM fits  and
the new values from deep inelastic scattering and $b\overline{b}$ lattice
calculations.}
\end{figure}
\subsection{\it MSSM Fits and Comparison with the SM}
%
%
As mentioned in the introduction, the MSSM can increase the value of $R_b$,
which experimentally is slightly above the SM value.
The major additional contributions originate from vertex contributions with light 
charginos and light right handed stops in the low $\tan\beta$ scenario
and light higgses for large $\tan\beta $ values. Since the large $\tan\beta$
scenario does not improve $R_b$ significantly\cite{hollik1}, it will not be discussed
here anymore.
 The $R_b$ dependence on  chargino and stop masses is shown in
fig. \ref{rb}.
 The experimental value $R_b=0.2178\pm 0.0011$ is clearly 
above the SM value of 0.2158 and can be obtained for   charginos around 
85 GeV  and stop masses around 50 GeV (best fit results).

The fit results  are 
compared with the Standard Model fits   in
fig.~\ref{\figV}. The  Standard Model
$\chi^2/d.o.f.=18.5/15$ corresponds to a pro\-bability of 24\%,
the MSSM  $\chi^2/d.o.f.=16.1/12$  to a probability of 19\%. 
In counting the d.o.f the insensitive (and fixed) parameters were ignored.

Another interesting point are the    $\alpha_s(M_Z)$ values.
An increase in $R_b$ implies in increase in the total width of the $Z^0$ boson, which
can be compensated by a decrease in the QCD corrections, i.e. $\alpha_s$.
Fig.~\ref{\alphas} shows a comparison of different measurements
of $\alpha_s(M_Z)$  with the fitted values given in this paper.
The agreement is remarkable.
Note that the $\alpha_s$ crisis has disappeared after the LEP value from the total
cross section came down and the value from both lattice calculations and deep 
inelastic scattering went up\cite{schmelling}.

\section{CMSSM and $R_b$}

An increase in $R_b$ requires
one (mainly right handed) stop to be light and the other one to be heavy.
If both would be light, then all other squarks are likely to be light, 
which would upset the good agreement between the SM 
and all other electroweak data.
A large mass splitting in the stop sector 
needs a very artificial fine tuning of the few free parameters in 
the Constrained MSSM, which assumes  unification of gauge and
b-$\tau$ Yukawa couplings\cite{cmssm}. 
This is obvious from the mixing matrix:
\begin{equation}
\small
 \label{stopmat}
\left(\begin{array}{cc}
\tilde{m}^2_{L}+m_t^2+D_L &
m_t(A_tm_0-\mu\cot \beta ) \\  
m_t(A_tm_0-\mu\cot \beta ) &   
\tilde{m}^2_{R}+m_t^2 +D_R
\end{array}  \right)       \nonumber
\end{equation}
\normalsize
The D-terms proportional to $\cos 2\beta$ are negligible 
for $\tan\beta\approx 1$.
If one of the diagonal elements is much larger than $m_t$, the off-diagonal
terms of the order  $m_t$ will not cause a mixing and 
the difference between the left- and 
right-handed stops has to come from the evolution of the diagonal terms: 
\begin{eqnarray}
\small
\frac{d\tilde{m}^2_{L}}{dt} & = & (\frac{16}{3}\tilde{\alpha}_3M^2_3
    + 3\tilde{\alpha}_2M^2_2 + \frac{1}{15}\tilde{\alpha}_1M^2_1)
      \label{RGEmQ} \\
 && - [Y_t(\tilde{m}^2_{Q_3}+\tilde{m}^2_{U_3}+m^2_{H_2}+A^2_t m_0^2)\nn
 &&    +Y_b(\tilde{m}^2_{Q_3}+\tilde{m}^2_{D_3}+m^2_{H_1}+A^2_b m_0^2)]\nn
\frac{d\tilde{m}^2_{R}}{dt} & = & (\frac{16}{3}\tilde{\alpha}_3M^2_3
    +\frac{16}{15}\tilde{\alpha}_1M^2_1)\\
&&    -2Y_t(\tilde{m}^2_{Q_3}+\tilde{m}^2_{U_3}  
 + m^2_{H_2}+A^2_t m_0^2)\nonumber \label{RGEmU}
\label{RGEmD}
\end{eqnarray}
\normalsize
One observes that
the difference between left- and right handed stops depends
on the Yukawa couplings for top and bottom ($Y_t,Y_b$) 
and the trilinear couplings $A_{t(b)}$. For low $\tan\beta$
$Y_b$ is negligible, while  $A_t$ and $Y_t$ are not free parameters, 
since they  go to 
 fixed point solutions\cite{cmssm}. Therefore there is no freedom to
adjust these parameters within the CMSSM in order to get a large splitting
between the left- and right-handed stops.  
\normalsize

In addition, problems arise with electroweak symmetry breaking,
since this requires the Higgs mixing parameter $\mu$ 
to be much heavier than the gaugino masses\cite{cmssm}, 
while $R_b$ requires low values of $\mu$ for a 
significant enhancement (since the
chargino has to be preferably Higgsino-like).
In conclusion, within the CMSSM an enhancement of $R_b$ above 
the SM is practically excluded.

\section{Conclusions}
Both the MSSM and SM provide a good description of all electroweak data.
With all mass bounds and the $b\rightarrow s\gamma$ rate 
included in the fit, the best
$\chi^2/d.o.f$ in the MSSM (SM) is  16.1/12 (18.5/15), which  corresponds to a
probability of 19\% (24\%).
The slightly better $\chi^2$ of the MSSM is mainly 
due to the better description 
of $R_b$, but this requires an unnatural large splitting in the stop sector.
  Final analysis of available LEP data  will teach of the 
present prelininary value of $R_b$ will indeed stay above the SM value.
%
\section{Acknowledgments}
Thanks go to the Electroweak Working Group and the LEP experiments, who made
the preliminary data available before the Conference and to my close 
collaborators 
Wolfgang Hollik and Ulrich Schwickerath during this analysis .
This work was done during a sabbatical and
support  from the Volkswagen-Stiftung (Contract I/71681)
is greatly appreciated.

\end{document}